# A quantum gravity tensor equation formally integrating general relativity with quantum mechanics

Xu Duan

*Zhengzhou City Construction Engineering Foundation Institute - 450000 Zhengzhou, People's Republic of China*

## Abstract

Extending black-hole entropy to ordinary objects, we propose kinetic entropy tensor, based on which a quantum gravity tensor equation is established. Our investigation results indicate that if $N=1$, the quantum gravity tensor equation returns to Schrödinger integral equation. When $N$ becomes sufficiently large, it is equivalent to Einstein field equation. This illustrates formal unification and intrinsic compatibility of general relativity with quantum mechanics. The quantum gravity equation may be utilized to deduce general relativity, special relativity, Newtonian mechanics and quantum mechanics, which has paved the way for unification of theoretical physics.

## 1. Introduction

The unification of general relativity and quantum mechanics, a long-standing unsolved difficult problem, is one of the most important frontier researches in theoretical physics. It has puzzled many famous grandmasters of science for decades, including Albert Einstein. A current viewpoint suggests that they are essentially contradictory in their ideas and explanation of our world. General relativity is a very effective theory concerning macroscopic objects, especially cosmology. But in contrast, quantum mechanics specializes in dealing with molecules and atoms. Both theories have their own scope of application, and seem to be incomplete. It appears that we can't simply combine these two theories to create a single equation of quantum gravity.

Although substantial exciting progresses, such as black hole entropy, holographic principle, M (superstring) theory and loop quantum gravity, have emerged in the domain of quantum gravity, yet up till now, all the attempts to discover a theory of formally unified and complete quantum gravity are far from succeeding.

Inspired by black-hole thermodynamics, we propose in this paper kinetic entropy tensor and successfully construct a quantum gravity tensor equation, which could integrate Schrödinger equation with Einstein field equation self-consistently. This means general relativity and quantum mechanics have intrinsic compatibility, instead of being irreconcilable with each other, as is thought previously. The quantum gravity equation may be utilized to deduce general relativity, special relativity, Newtonian mechanics and quantum mechanics, which has paved the way for unification of theoretical physics.

E-mail: duanxu760613@stu.huel.edu.cn

## 2. Derivation of kinetic entropy

In 1976 William G. Unruh[1] put forward his famous Unruh effect, indicating that accelerated motion is equivalent to thermal radiation of black bodies. It gives a strong hint that temperature and entropy are closely related to kinetics. However, previously the concept of entropy was only restricted to classical equilibrium state in thermodynamics. To deduce kinetic entropy, we need to take advantage of entropy force together with Unruh formula.

Entropy force[2,3], as a real force driven by entropy changes of a system, could be written as follows:

$$TdS = FdR \qquad (1)$$

among which, $T$ and $S$ represent temperature and entropy, respectively. $F$ is entropy force and $R$ is space coordinate.

Unruh law[1] in quantum field theory is expressed by

$$kT = \frac{1}{2\pi} \cdot \frac{\hbar a}{c} \qquad (2)$$

where $T$ is Unruh temperature and $a$ is acceleration. $k$, $\hbar$ and $c$ are respectively Boltzmann constant, reduced Planck constant and velocity of light in vacuum.

By combining (1) with (2), we have:

$$Mc^2 \cdot kT \cdot dR = \frac{\hbar c}{2\pi} TdS \qquad (3)$$

Furthermore, we acquire:

$$dS = 2\pi \frac{k}{\hbar c}(Mc^2) dR \qquad (4)$$

The integral form of kinetic entropy is:

$$\Delta S = 2\pi \frac{k}{\hbar c}(Mc^2) R \qquad (5)$$

## 3. Kinetic entropy and Unruh temperature of general objects

Black-hole thermodynamics developed during 1970's reveals that black holes possess temperature and entropy. In fact, general objects (not only black holes) also have kinetic entropy and Unruh temperature. So how to extend thermodynamics of black holes to general objects?

First we introduce escape velocity $v = \sqrt{2GM/R}$. It is not hard to find that energy $U$、space $R$ and escape velocity $v$ could be generalized into the following relationship, which applies to an ordinary object.

$$\frac{U}{E_p} = \frac{R}{x_p}\left(\frac{v^2}{2c^2}\right) \qquad (6)$$

$x_p$ called Planck scale, $x_P = \sqrt{G\hbar/c^3}$, $E_P$ called Planck energy, $E_P = \sqrt{\hbar c^5/G}$, $U = Mc^2$, G representing universal gravitational constant.

In appendix (the last section), we will see that (6) could be derived directly from Friedmann equation in flat space-time. Based on (6), kinetic entropy and Unruh temperature of ordinary objects are easy to obtain, including black-hole entropy and black-hole temperature.

Specifically speaking, putting (6) into (5), we obtain kinetic entropy of ordinary objects:

$$\Delta S = 2\pi k \cdot \frac{U}{E_p} \cdot \frac{R}{x_p} = 2\pi k \left(\frac{R}{x_p}\right)^2 \left(\frac{v^2}{2c^2}\right) \tag{7}$$

If $v = c$, it's just Bekenstein's black-hole entropy[4].

$$\Delta S_{BH} = \pi k \left(\frac{R}{x_p}\right)^2 = \pi k \frac{R^2 c^3}{G\hbar} \tag{8}$$

Furthermore, by transformation of Unruh formula, we get:

$$kT \cdot R = \frac{\hbar c}{2\pi}\left(\frac{v^2}{2c^2}\right) \tag{9}$$

After (9) is plugged into (6), Unruh temperature of general objects is acquired:

$$kT = \frac{E_p^2}{Mc^2} \cdot \frac{1}{2\pi}\left(\frac{v^2}{2c^2}\right)^2 \tag{10}$$

In particular, when $v = c$, black-hole temperature derived by Hawking [5] is easily got from our method:

$$kT = \frac{E_p^2}{8\pi Mc^2} = \frac{\hbar c^3}{8\pi GM}$$

It is not hard to prove that kinetic entropy and Unruh temperature of general objects invariably satisfy the relationship below, which is very useful in the deduction of Newton's law later.

$$T\Delta S = \frac{1}{2}Mv^2 \tag{11}$$

## 4. Kinetic entropy tensor

Maybe an important enlightenment of general relativity is that theories in the form of tensors are genuinely complete. The key to unification of general relativity and quantum mechanics lies in that kinetic entropy must be rewritten as a second-order symmetrical tensor.

How to determine the expressions of kinetic entropy tensor $S_{ab}$? Obviously by analogy with kinetic entropy (5), we presume kinetic entropy tensor $S_{ab}$ should be defined as:

$$S_{ab} \stackrel{def}{=} 2\pi \frac{k}{\hbar c} U_{ab} \cdot I_{ab} \cdot X_{ab} \tag{12}$$

Here $U_{ab} = T_{ab} \cdot \left(\frac{4}{3}\pi r^3\right)$, and $T_{ab}$ denotes energy-momentum tensor. $X_{ab}$ stands for tensor of space-time coordinate, and $I_{ab}$ is coefficient matrix. The next crucial step is to write out the specific components of $U_{ab}$ and $X_{ab}$ under Cartesian coordinate system, although they are arbitrary and depend on selection of metric tensor. But we know $U_{ab}$ has a standard form:

$$U_{ab} = \begin{bmatrix} Mc^2 & P_x c & P_y c & P_z c \\ P_x c & Mv_x^2 & Mv_x v_y & Mv_x v_z \\ P_y c & Mv_y v_x & Mv_y^2 & Mv_y v_z \\ P_z c & Mv_z v_x & Mv_z v_y & Mv_z^2 \end{bmatrix} = Mc^2 \cdot \begin{bmatrix} 1 & \frac{v_x}{c} & \frac{v_y}{c} & \frac{v_z}{c} \\ \frac{v_x}{c} & \frac{v_x^2}{c^2} & \frac{v_x v_y}{c^2} & \frac{v_x v_z}{c^2} \\ \frac{v_y}{c} & \frac{v_x v_y}{c^2} & \frac{v_y^2}{c^2} & \frac{v_y v_z}{c^2} \\ \frac{v_z}{c} & \frac{v_x v_z}{c^2} & \frac{v_y v_z}{c^2} & \frac{v_z^2}{c^2} \end{bmatrix} = Mc^2 \cdot P_{ab}$$

To satisfy the covariant principle, we designate $X_{ab}$ adopt the same symmetrical form:

$$X_{ab} = \begin{bmatrix} ct & x & y & z \\ x & x \cdot \frac{v_x}{c} & x \cdot \frac{v_y}{c} & x \cdot \frac{v_z}{c} \\ y & y \cdot \frac{v_x}{c} & y \cdot \frac{v_y}{c} & y \cdot \frac{v_z}{c} \\ z & z \cdot \frac{v_x}{c} & z \cdot \frac{v_y}{c} & z \cdot \frac{v_z}{c} \end{bmatrix} = ct \cdot \begin{bmatrix} 1 & \frac{v_x}{c} & \frac{v_y}{c} & \frac{v_z}{c} \\ \frac{v_x}{c} & \frac{v_x^2}{c^2} & \frac{v_x v_y}{c^2} & \frac{v_x v_z}{c^2} \\ \frac{v_y}{c} & \frac{v_x v_y}{c^2} & \frac{v_y^2}{c^2} & \frac{v_y v_z}{c^2} \\ \frac{v_z}{c} & \frac{v_x v_z}{c^2} & \frac{v_y v_z}{c^2} & \frac{v_z^2}{c^2} \end{bmatrix} = ct \cdot P_{ab}$$

After repeated speculations and tentative calculations, an important matrix relation is ultimately obtained:

$$\begin{bmatrix} 1 & \frac{v_x}{c} & \frac{v_y}{c} & \frac{v_z}{c} \\ \frac{v_x}{c} & \frac{v_x^2}{c^2} & \frac{v_x v_y}{c^2} & \frac{v_x v_z}{c^2} \\ \frac{v_y}{c} & \frac{v_x v_y}{c^2} & \frac{v_y^2}{c^2} & \frac{v_y v_z}{c^2} \\ \frac{v_z}{c} & \frac{v_x v_z}{c^2} & \frac{v_y v_z}{c^2} & \frac{v_z^2}{c^2} \end{bmatrix} \cdot \begin{bmatrix} -\frac{v^2}{2c^2} & 0 & 0 & 0 \\ 0 & 1 & 0 & 0 \\ 0 & 0 & 1 & 0 \\ 0 & 0 & 0 & 1 \end{bmatrix} \cdot \begin{bmatrix} 1 & \frac{v_x}{c} & \frac{v_y}{c} & \frac{v_z}{c} \\ \frac{v_x}{c} & \frac{v_x^2}{c^2} & \frac{v_x v_y}{c^2} & \frac{v_x v_z}{c^2} \\ \frac{v_y}{c} & \frac{v_x v_y}{c^2} & \frac{v_y^2}{c^2} & \frac{v_y v_z}{c^2} \\ \frac{v_z}{c} & \frac{v_x v_z}{c^2} & \frac{v_y v_z}{c^2} & \frac{v_z^2}{c^2} \end{bmatrix}$$

$$= \frac{v^2}{2c^2} \cdot \begin{bmatrix} 1 & \frac{v_x}{c} & \frac{v_y}{c} & \frac{v_z}{c} \\ \frac{v_x}{c} & \frac{v_x^2}{c^2} & \frac{v_x v_y}{c^2} & \frac{v_x v_z}{c^2} \\ \frac{v_y}{c} & \frac{v_x v_y}{c^2} & \frac{v_y^2}{c^2} & \frac{v_y v_z}{c^2} \\ \frac{v_z}{c} & \frac{v_x v_z}{c^2} & \frac{v_y v_z}{c^2} & \frac{v_z^2}{c^2} \end{bmatrix}$$

So we learn proper choice of coefficient matrix $I_{ab}$ helps to make $S_{ab}$, $U_{ab}$ and $X_{ab}$ always covariant. Hence coefficient matrix $I_{ab}$ is determined as follows:

$$P_{ab} \cdot I_{ab} \cdot P_{ab} = \left(\frac{v^2}{2c^2}\right) \cdot P_{ab}$$

Or

$$I_{ab} = \begin{bmatrix} -\dfrac{v^2}{2c^2} & 0 & 0 & 0 \\ 0 & 1 & 0 & 0 \\ 0 & 0 & 1 & 0 \\ 0 & 0 & 0 & 1 \end{bmatrix}_{4\times 4}$$

If we employ Bits number (or degree of freedom) tensor $N_{ab}$ to fulfill the quantization of entropy:

$$S_{ab} = N_{ab} k$$

then there is:

$$2\pi U_{ab} \cdot I_{ab} \cdot X_{ab} = N_{ab} \hbar c \tag{13}$$

We have actually established a half of the quantum gravity equation.

## 5. Wave function tensor and kinetic entropy tensor

Is there any correlation between kinetic entropy and wave function? Our investigation offers a definite answer. Seeing that kinetic entropy is quantized, we have:

$$\Delta S = 2\pi \frac{k}{\hbar c} Mc^2 \cdot r = N' \cdot k \quad (\text{$N'$ is Bits number or degree of freedom})$$

$$N' = 2\pi \frac{Mc^2 \cdot r}{\hbar c}$$

Supposing a macroscopic body is made up of $N$ microscopic particles. $M$ means mass of a macroscopic body, and $m$ refers to mass of a microscopic particle. It is obvious $M = N \cdot m$.

In addition, wave function obeys Schrödinger equation:

$$\psi = \psi_0 \exp\left\{-\frac{m}{2i\hbar} \cdot \frac{r^2}{t}\right\}$$

Thus

$$-2\pi N \cdot i \ln\left(\frac{\psi}{\psi_0}\right)^2 = 2\pi \frac{N \cdot m}{\hbar} \cdot \frac{r^2}{t} = 2\pi \frac{Mc^2}{\hbar c} \cdot r \cdot \frac{v}{c}$$

$$= N' \cdot \frac{v}{c}$$

This implies Bits number (or degree of freedom) $N'$ and particle number $N$ should satisfy the relation:

$$N' \cdot \left(\frac{v}{c}\right) = -2\pi \cdot N \cdot i \ln\left(\frac{\psi}{\psi_0}\right)^2 \tag{14}$$

In order to fit kinetic entropy tensor and Bits number tensor, we have to extend wave function to a second-order symmetrical tensor $\psi_{ab}$. When we define wave function tensor as:

$$\ln \psi_{ab} \stackrel{\text{def}}{=} \left( \ln \frac{\psi}{\psi_0} \right) \cdot \begin{bmatrix} 1 & \frac{v_x}{c} & \frac{v_y}{c} & \frac{v_z}{c} \\ \frac{v_x}{c} & \frac{v_x^2}{c^2} & \frac{v_x v_y}{c^2} & \frac{v_x v_z}{c^2} \\ \frac{v_y}{c} & \frac{v_x v_y}{c^2} & \frac{v_y^2}{c^2} & \frac{v_y v_z}{c^2} \\ \frac{v_z}{c} & \frac{v_x v_z}{c^2} & \frac{v_y v_z}{c^2} & \frac{v_z^2}{c^2} \end{bmatrix}_{4 \times 4}$$

we discover there exists simple correspondence between wave function tensor $\psi_{ab}$ and Bits number tensor $N_{ab}$ via a series of matrix calculations:

$$N_{ab} = -2\pi \cdot N \cdot i \ln \psi_{ab}$$

Detailed calculations are given as following:

$$N_{ab} = \frac{2\pi U_{ab} \cdot I_{ab} \cdot X_{ab}}{\hbar c}$$

$$= 2\pi \frac{Mc^2}{E_p} \cdot \frac{ct}{x_p} \begin{bmatrix} 1 & \frac{v_x}{c} & \frac{v_y}{c} & \frac{v_z}{c} \\ \frac{v_x}{c} & \frac{v_x^2}{c^2} & \frac{v_x v_y}{c^2} & \frac{v_x v_z}{c^2} \\ \frac{v_y}{c} & \frac{v_x v_y}{c^2} & \frac{v_y^2}{c^2} & \frac{v_y v_z}{c^2} \\ \frac{v_z}{c} & \frac{v_x v_z}{c^2} & \frac{v_y v_z}{c^2} & \frac{v_z^2}{c^2} \end{bmatrix} \cdot \begin{bmatrix} -\frac{v^2}{2c^2} & 0 & 0 & 0 \\ 0 & 1 & 0 & 0 \\ 0 & 0 & 1 & 0 \\ 0 & 0 & 0 & 1 \end{bmatrix} \cdot \begin{bmatrix} 1 & \frac{v_x}{c} & \frac{v_y}{c} & \frac{v_z}{c} \\ \frac{v_x}{c} & \frac{v_x^2}{c^2} & \frac{v_x v_y}{c^2} & \frac{v_x v_z}{c^2} \\ \frac{v_y}{c} & \frac{v_x v_y}{c^2} & \frac{v_y^2}{c^2} & \frac{v_y v_z}{c^2} \\ \frac{v_z}{c} & \frac{v_x v_z}{c^2} & \frac{v_y v_z}{c^2} & \frac{v_z^2}{c^2} \end{bmatrix}$$

$$= 2\pi \frac{Mc^2}{E_p} \cdot \frac{R}{x_p} \cdot \frac{v}{2c} \begin{bmatrix} 1 & \frac{v_x}{c} & \frac{v_y}{c} & \frac{v_z}{c} \\ \frac{v_x}{c} & \frac{v_x^2}{c^2} & \frac{v_x v_y}{c^2} & \frac{v_x v_z}{c^2} \\ \frac{v_y}{c} & \frac{v_x v_y}{c^2} & \frac{v_y^2}{c^2} & \frac{v_y v_z}{c^2} \\ \frac{v_z}{c} & \frac{v_x v_z}{c^2} & \frac{v_y v_z}{c^2} & \frac{v_z^2}{c^2} \end{bmatrix} = 2\pi \cdot \left( \frac{R}{x_p} \right)^2 \cdot \frac{v^3}{4c^3} \begin{bmatrix} 1 & \frac{v_x}{c} & \frac{v_y}{c} & \frac{v_z}{c} \\ \frac{v_x}{c} & \frac{v_x^2}{c^2} & \frac{v_x v_y}{c^2} & \frac{v_x v_z}{c^2} \\ \frac{v_y}{c} & \frac{v_x v_y}{c^2} & \frac{v_y^2}{c^2} & \frac{v_y v_z}{c^2} \\ \frac{v_z}{c} & \frac{v_x v_z}{c^2} & \frac{v_y v_z}{c^2} & \frac{v_z^2}{c^2} \end{bmatrix}$$

$$= \frac{1}{2} N' \cdot \left( \frac{v}{c} \right) \cdot \begin{bmatrix} 1 & \frac{v_x}{c} & \frac{v_y}{c} & \frac{v_z}{c} \\ \frac{v_x}{c} & \frac{v_x^2}{c^2} & \frac{v_x v_y}{c^2} & \frac{v_x v_z}{c^2} \\ \frac{v_y}{c} & \frac{v_x v_y}{c^2} & \frac{v_y^2}{c^2} & \frac{v_y v_z}{c^2} \\ \frac{v_z}{c} & \frac{v_x v_z}{c^2} & \frac{v_y v_z}{c^2} & \frac{v_z^2}{c^2} \end{bmatrix}$$

$$= -2\pi \cdot N \cdot i \ln \psi_{ab}$$

This leads to a complete quantum gravity tensor equation:

$$2\pi U_{ab} \cdot I_{ab} \cdot X_{ab} = N_{ab} \hbar c = -2\pi \cdot N \hbar c \cdot i \ln \psi_{ab} \tag{15}$$

## 6. Discussions about quantum gravity equation

In this section we will focus on particle number $N$. Two special cases will be discussed in terms of $N$ value.

When $N=1$, (corresponding to a single free particle), quantum gravity equation (15) becomes:

$$mc^2 \cdot \begin{bmatrix} 1 & \frac{v_x}{c} & \frac{v_y}{c} & \frac{v_z}{c} \\ \frac{v_x}{c} & \frac{v_x^2}{c^2} & \frac{v_x v_y}{c^2} & \frac{v_x v_z}{c^2} \\ \frac{v_y}{c} & \frac{v_x v_y}{c^2} & \frac{v_y^2}{c^2} & \frac{v_y v_z}{c^2} \\ \frac{v_z}{c} & \frac{v_x v_z}{c^2} & \frac{v_y v_z}{c^2} & \frac{v_z^2}{c^2} \end{bmatrix} \cdot \begin{bmatrix} -\frac{v^2}{2c^2} & 0 & 0 & 0 \\ 0 & 1 & 0 & 0 \\ 0 & 0 & 1 & 0 \\ 0 & 0 & 0 & 1 \end{bmatrix} \cdot ct \cdot \begin{bmatrix} 1 & \frac{v_x}{c} & \frac{v_y}{c} & \frac{v_z}{c} \\ \frac{v_x}{c} & \frac{v_x^2}{c^2} & \frac{v_x v_y}{c^2} & \frac{v_x v_z}{c^2} \\ \frac{v_y}{c} & \frac{v_x v_y}{c^2} & \frac{v_y^2}{c^2} & \frac{v_y v_z}{c^2} \\ \frac{v_z}{c} & \frac{v_x v_z}{c^2} & \frac{v_y v_z}{c^2} & \frac{v_z^2}{c^2} \end{bmatrix}$$

$$= -i\hbar c \cdot \left( \ln \frac{\psi}{\psi_0} \right) \cdot \begin{bmatrix} 1 & \frac{v_x}{c} & \frac{v_y}{c} & \frac{v_z}{c} \\ \frac{v_x}{c} & \frac{v_x^2}{c^2} & \frac{v_x v_y}{c^2} & \frac{v_x v_z}{c^2} \\ \frac{v_y}{c} & \frac{v_x v_y}{c^2} & \frac{v_y^2}{c^2} & \frac{v_y v_z}{c^2} \\ \frac{v_z}{c} & \frac{v_x v_z}{c^2} & \frac{v_y v_z}{c^2} & \frac{v_z^2}{c^2} \end{bmatrix}$$

By detailed matrix calculations, Schrödinger equation of a single free particle emerges exactly in integral form:

$$p_x \cdot x + p_y \cdot y + p_z \cdot z - \varepsilon t = -i\hbar \ln \frac{\psi}{\psi_0} \qquad \varepsilon = \frac{1}{2}mv^2$$

Of course, here we assume at initial time $t_0=0$, the particle is located at the origin of coordinates (initial coordinate $r_0=0$). It further leads to the inference:

$$\psi = \psi_0 \exp\left\{ -\frac{m}{2i\hbar} \cdot \frac{r^2}{t} \right\}$$

That is to say, Schrödinger equation in integral form could be derived from quantum gravity equation (15) with $N=1$.

While $N$ is sufficiently large (for example, $N \cong 10^{23}$), our research subject is converted to a macroscopic object containing enough microscopic particles. Here three basic principles must be satisfied simultaneously:

(1) The macroscopic object obeys Newtonian mechanics:

$$v = \sqrt{\frac{2GM}{R}}$$

(2) Every microscopic particle within the macroscopic object obeys Schrödinger equation:

$$\psi = \psi_0 \exp\left\{ -\frac{m}{2i\hbar} \cdot \frac{r^2}{t} \right\}$$

(3) Holographic principle

The following calculation manifests that Bits number tensor $N_{ab}$ can only equal to:

$$N_{ab} = 2\pi \cdot \frac{X_{ab}}{x_p} \cdot I_{ab} \cdot \frac{X_{ab}}{x_p} \cdot \frac{v^3}{2c^3}$$

Details are given below:

$$N_{ab} = -2\pi \cdot N \cdot i \ln \psi_{ab} = -2\pi \cdot N \cdot i \ln \frac{\psi}{\psi_0} \cdot \begin{bmatrix} 1 & \frac{v_x}{c} & \frac{v_y}{c} & \frac{v_z}{c} \\ \frac{v_x}{c} & \frac{v_x^2}{c^2} & \frac{v_x v_y}{c^2} & \frac{v_x v_z}{c^2} \\ \frac{v_y}{c} & \frac{v_x v_y}{c^2} & \frac{v_y^2}{c^2} & \frac{v_y v_z}{c^2} \\ \frac{v_z}{c} & \frac{v_x v_z}{c^2} & \frac{v_y v_z}{c^2} & \frac{v_z^2}{c^2} \end{bmatrix}$$

$$= -2\pi \cdot N \cdot i \ln \frac{\psi}{\psi_0} \cdot P_{ab}$$

$$= 2\pi \cdot N \cdot \frac{m}{2\hbar} \cdot \frac{r^2}{t} \cdot P_{ab} = 2\pi \cdot \frac{M}{2\hbar} \cdot \frac{r^2}{t} \cdot P_{ab} = 2\pi \cdot \frac{2GM}{r} \cdot \frac{r^2 v}{4G\hbar} \cdot P_{ab} = 2\pi \cdot \frac{r^2 \cdot v^3}{4G\hbar} \cdot P_{ab}$$

$$= 2\pi \cdot \left(\frac{r}{x_p}\right)^2 \cdot \frac{v^3}{4c^3} \cdot P_{ab} = 2\pi \cdot \left(\frac{ct}{x_p}\right)^2 \cdot \frac{v^3}{2c^3} \cdot P_{ab} \cdot I_{ab} \cdot P_{ab}$$

$$= 2\pi \cdot \frac{X_{ab}}{x_p} \cdot I_{ab} \cdot \frac{X_{ab}}{x_p} \cdot \frac{v^3}{2c^3}$$

It means that quantum gravity equation (15) with N sufficiently large is practically equivalent to:

$$2\pi \cdot \frac{U_{ab}}{E_p} \cdot I_{ab} \cdot \frac{X_{ab}}{x_p} = N_{ab} = 2\pi \cdot \frac{X_{ab}}{x_p} \cdot I_{ab} \cdot \frac{X_{ab}}{x_p} \cdot \frac{v^3}{2c^3}$$

Quite evidently, we have:

$$\frac{U_{ab}}{E_p} = \frac{X_{ab}}{x_p} \left(\frac{v^3}{2c^3}\right)$$

Provided both sides of the above are divided by $V = (4/3)\pi r^3$, we obtain at once:

$$\frac{8\pi G}{c^4} \rho c^2 \cdot \begin{bmatrix} 1 & \frac{v_x}{c} & \frac{v_y}{c} & \frac{v_z}{c} \\ \frac{v_x}{c} & \frac{v_x^2}{c^2} & \frac{v_x v_y}{c^2} & \frac{v_x v_z}{c^2} \\ \frac{v_y}{c} & \frac{v_x v_y}{c^2} & \frac{v_y^2}{c^2} & \frac{v_y v_z}{c^2} \\ \frac{v_z}{c} & \frac{v_x v_z}{c^2} & \frac{v_y v_z}{c^2} & \frac{v_z^2}{c^2} \end{bmatrix} = \left(\frac{3v^2}{r^2 c^2}\right) \cdot \begin{bmatrix} 1 & \frac{v_x}{c} & \frac{v_y}{c} & \frac{v_z}{c} \\ \frac{v_x}{c} & \frac{v_x^2}{c^2} & \frac{v_x v_y}{c^2} & \frac{v_x v_z}{c^2} \\ \frac{v_y}{c} & \frac{v_x v_y}{c^2} & \frac{v_y^2}{c^2} & \frac{v_y v_z}{c^2} \\ \frac{v_z}{c} & \frac{v_x v_z}{c^2} & \frac{v_y v_z}{c^2} & \frac{v_z^2}{c^2} \end{bmatrix}$$

According to appendix, this is precisely Einstein field equation under RW metric in flat space-time:

$$\frac{8\pi G}{c^4} T_{ab} = G_{ab}$$

Interestingly, for an object, when it contains only one free particle ( N=1 ), it is subject to Schrödinger equation. When it includes enough microscopic particles (we call it a macroscopic body), thanks to domination of holographic principle[2], it conforms to general relativity. This explains the reason why Einstein field equation only applies to macroscopic bodies, instead of microscopic particles.

It should be stressed that theoretical physicists mistakenly believe there exists incompatibility between general relativity and quantum mechanics. But our research demonstrates that at least under RW metric, Einstein field equation and Schrödinger equation could be unified. Besides, our perspective is quite different from superstring and loop quantum gravity.

Now let's return again to kinetic entropy (5) and Unruh formula (2) to reproduce Newton's law. Referring to Verlinde's paper[2], just putting (5) and (2) into (11), we get Newton's second law $F = Ma$ immediately.

In addition, if we put (7) into (11), with Unruh formula (2), Newton's law of universal gravitation is also successfully duplicated:

$$a = \frac{GM}{R^2} \quad \text{and} \quad F = M'a = \frac{GMM'}{R^2}$$

## 7. The generalized quantum gravity equation

The quantum gravity equation (15) discussed in the 6th section is based on RW metric in flat space-time. We should generalize (15) so that it is tenable for arbitrary metrics in curved space-time. We have known, under RW metric in flat space-time,

$$G_{ab} = \left(\frac{3v^2}{r^2 c^2}\right) \cdot P_{ab} \quad \text{and} \quad X_{ab} = (ct) \cdot P_{ab}$$

So

$$X_{ab} = G_{ab} \cdot \frac{1}{3}(ct)^3$$

Plus

$$U_{ab} = T_{ab} \cdot \left(\frac{4}{3}\pi r^3\right)$$

Substitution of them into (15), the generalized quantum gravity equation is obtained:

$$\frac{4}{9}\pi r^3 \cdot (ct)^3 \cdot T_{ab} \cdot I_{ab} \cdot G_{ab} = \frac{N_{ab} \cdot \hbar c}{2\pi} = -N\hbar c \cdot i \ln \psi_{ab} \qquad (16)$$

in which $G_{ab}$ means Einstein tensor, and $T_{ab}$ means energy-momentum tensor. Coefficient matrix $I_{ab}$ must satisfy:

$$T_{ab} \cdot I_{ab} \cdot T_{ab} = \left(\frac{\rho}{2}v^2\right) \cdot T_{ab}$$

Considering N=1, under RW metric in flat space-time, ( $\kappa = 0$ ), as long as we write down:

$$T_{ab} = (\rho c^2) \cdot P_{ab}$$

$$G_{ab} = \left(\frac{3v^2}{r^2 c^2}\right) \cdot P_{ab}$$

$$\ln \psi_{ab} = \left(\ln \frac{\psi}{\psi_0}\right) \cdot P_{ab}$$

$$P_{ab} \cdot I_{ab} \cdot P_{ab} = \left(\frac{v^2}{2c^2}\right) \cdot P_{ab}$$

the direct inference of Schrödinger equation will be reproduced:

$$\psi = \psi_0 \exp\left\{-\frac{m}{2i\hbar} \cdot \frac{r^2}{t}\right\}$$

It manifests that generalized quantum gravity equation (16) might return to Schrödinger equation spontaneously under RW metric in flat space-time. Thereby we can draw a conclusion that Schrödinger equation holds for flat space-time background which is isotropic and even.

While N is sufficiently large, holographic principle begins to play a leading role. For arbitrary metrics, holographic principle takes the form of:

$$N_{ab} = 2\pi \cdot \frac{X_{ab}}{x_p} \cdot I_{ab} \cdot \frac{X_{ab}}{x_p} \cdot \frac{v^3}{2c^3}$$

$$= \frac{\pi}{9} r^3 \cdot (ct)^3 \cdot \frac{G_{ab}}{x_p} \cdot I_{ab} \cdot \frac{G_{ab}}{x_p}$$

If we plug it into generalized quantum gravity equation (16), Einstein field equation arises immediately:

$$\frac{8\pi G}{c^4} T_{ab} = G_{ab}$$

It signifies, by means of holographic principle, equation (16) could accurately return to Einstein field equation. In fact, it is holographic principle that makes probability give way to certainty for macroscopic bodies.

## 8. Special relativity and kinetic entropy

Finally we tentatively deduce special relativity only from kinetic entropy. The evident superiority of kinetic entropy lies in its simple and convenient deduction of special relativity, without use of Lorentz transformation and the hypothesis of 'constancy of light velocity'. We know (7) gives:

$$\Delta S = \pi k \left(\frac{R}{x_p}\right)^2 \cdot \frac{v^2}{c^2}$$

If we regard black hole entropy as the maximum equilibrium entropy $S_0$, which equals

$$S_0 = \pi k \left(\frac{R}{x_p}\right)^2$$

then non-equilibrium entropy $S$ can be represented as:

$$S = S_0 - \Delta S = S_0\left(1 - \frac{v^2}{c^2}\right)$$

In practice we have got Lorentz relationship of entropy with velocity. It is amazing that other physical quantities varying with velocity are all derived from it very easily.

(1) For instance, we derive the relation of space interval $R$ with velocity. As we know:

$$\Delta S = 2\pi k \left(\frac{R}{x_p}\right)^2 \left(\frac{v^2}{2c^2}\right)$$

Meanwhile, covariance requires:

$$\Delta S_0 = 2\pi k \left(\frac{R_0}{x_p}\right)^2 \left(\frac{v^2}{2c^2}\right)$$

We conclude immediately:

$$R = R_0\sqrt{1 - \frac{v^2}{c^2}}$$

(2) Next, we attempt to derive the relation of particle number $N$ with velocity. Quantization of entropy is taken into consideration:

$$S = Nk$$

Of course, covariance requires:

$$S_0 = N_0 k$$

Thus Lorentz relationship of $N$ with velocity is easily got:

$$N = N_0\left(1 - \frac{v^2}{c^2}\right)$$

(3) The following equations are needed to develop the relation of macroscopic mass $M$ with velocity:

$$\frac{Mc^2}{E_p} = \frac{R}{x_p}\left(\frac{v^2}{2c^2}\right)$$

$$\frac{M_0 c^2}{E_p} = \frac{R_0}{x_p}\left(\frac{v^2}{2c^2}\right)$$

Obviously we could reach a conclusion:

$$M = M_0\sqrt{1 - \frac{v^2}{c^2}}$$

(4) The relation of microscopic mass $m$ with velocity is easily derived on the basis of:

$$M = Nm$$

$$M_0 = N_0 m_0$$

Synthesizing the above conclusions, we have:

$$m = \frac{m_0}{\sqrt{1-\frac{v^2}{c^2}}}$$

(5) We further explore the relation of temperature $T$ with velocity from equipartition theorem:

$$M = 2\int_S TdN$$

$$M_0 = 2\int_S T_0 dN_0$$

It's not hard to infer:

$$T = \frac{T_0}{\sqrt{1-\frac{v^2}{c^2}}}$$

Maybe Lorentz transformation and principle of constancy of light velocity are both unnecessary, only kinetic entropy will suffice.

*************************************************************

## Appendix

### RW metric and matrix components of Einstein field equation

First of all, two rational hypotheses are given: (1) Space time is inherently characterized by cosmological principle and Robertson-Walker metric[6], which suggest space is isotropic and matter is evenly-distributed. (2) Space is flat, i.e. curvature constant in RW metric $\kappa \equiv 0$. So RW metric is simplified into:

$$ds^2 = -c^2 dt^2 + R^2(t)\left[dr^2 + r^2(d\theta^2 + \sin^2\theta d\varphi^2)\right]$$

Complicated calculations show that under RW metric with $\kappa = 0$, the nonzero components of Ricci tensor are: (here space coordinate is converted into $r$, so as to distinguish it from Ricci scalar)

$$R_{00} = -3\frac{a}{rc^2}$$

$$R_{ab} = -\frac{1}{c^2}\left(\frac{a}{r} + 2\frac{v^2}{r^2}\right)g_{ab} \quad (\text{a,b}=1,2,3)$$

Ricci scalar is

$$R = -\frac{6}{c^2}\left(\frac{a}{r} + \frac{v^2}{r^2}\right)$$

So the relations between Einstein tensor and metric tensor are gained:

$$G_{00} = 3\frac{v^2}{r^2c^2}$$

$$G_{ab} = \left(2\frac{a}{rc^2} + \frac{v^2}{r^2c^2}\right)g_{ab} \quad (a,b=1,2,3)$$

Just demand

$$\lambda = -\left(2\frac{a}{rc^2} + \frac{v^2}{r^2c^2}\right)\bigg/3\frac{v^2}{r^2c^2}$$

Einstein tensor is written as below:

$$G_{ab} = \left(3\frac{v^2}{r^2c^2}\right)\cdot\begin{bmatrix} 1 & 0 & 0 & 0 \\ 0 & -\lambda & 0 & 0 \\ 0 & 0 & -\lambda & 0 \\ 0 & 0 & 0 & -\lambda \end{bmatrix}$$

As we know, energy-momentum tensor conforming to cosmological principle[7] gives:

$$T_{ab} = (\rho + \frac{p}{c^2})\, v_a v_b - p\cdot g_{ab}$$

The nonzero components are listed as:

$$T_{00} = \rho c^2$$

$$T_{ab} = -\lambda\cdot(\rho c^2)\cdot g_{ab} \quad (a,b=1,2,3)$$

We have accordingly:

$$T_{ab} = (\rho c^2)\cdot\begin{bmatrix} 1 & 0 & 0 & 0 \\ 0 & -\lambda & 0 & 0 \\ 0 & 0 & -\lambda & 0 \\ 0 & 0 & 0 & -\lambda \end{bmatrix}$$

Consequently, matrix components of Einstein field equation are represented as:

$$\frac{8\pi G}{c^4}\rho c^2 \cdot \begin{bmatrix} 1 & 0 & 0 & 0 \\ 0 & -\lambda & 0 & 0 \\ 0 & 0 & -\lambda & 0 \\ 0 & 0 & 0 & -\lambda \end{bmatrix} = \left(\frac{3v^2}{r^2c^2}\right)\cdot\begin{bmatrix} 1 & 0 & 0 & 0 \\ 0 & -\lambda & 0 & 0 \\ 0 & 0 & -\lambda & 0 \\ 0 & 0 & 0 & -\lambda \end{bmatrix}$$

It directly results in Friedmann equation ($\kappa = 0$):

$$8\pi\frac{\rho c^2}{E_p} = \frac{1}{x_p}\left(\frac{3v^2}{r^2c^2}\right)$$

Provided that both sides of Friedmann equation are multiplied by $V = \frac{4}{3}\pi r^3$, we get (6):

$$\frac{Mc^2}{E_p} = \frac{r}{x_p}\left(\frac{v^2}{2c^2}\right)$$

Consider a body in motion relative to coordinate system. Here co-moving coordinates can't be adopted. So energy-momentum tensor is given by

$$T_{ab} = (\rho c^2) \cdot \begin{bmatrix} 1 & \frac{v_x}{c} & \frac{v_y}{c} & \frac{v_z}{c} \\ \frac{v_x}{c} & \frac{v_x^2}{c^2} & \frac{v_x v_y}{c^2} & \frac{v_x v_z}{c^2} \\ \frac{v_y}{c} & \frac{v_x v_y}{c^2} & \frac{v_y^2}{c^2} & \frac{v_y v_z}{c^2} \\ \frac{v_z}{c} & \frac{v_x v_z}{c^2} & \frac{v_y v_z}{c^2} & \frac{v_z^2}{c^2} \end{bmatrix}$$

One may write down the components (matrix form) of field equation at this point:

$$\frac{8\pi G}{c^4}\rho c^2 \cdot \begin{bmatrix} 1 & \frac{v_x}{c} & \frac{v_y}{c} & \frac{v_z}{c} \\ \frac{v_x}{c} & \frac{v_x^2}{c^2} & \frac{v_x v_y}{c^2} & \frac{v_x v_z}{c^2} \\ \frac{v_y}{c} & \frac{v_x v_y}{c^2} & \frac{v_y^2}{c^2} & \frac{v_y v_z}{c^2} \\ \frac{v_z}{c} & \frac{v_x v_z}{c^2} & \frac{v_y v_z}{c^2} & \frac{v_z^2}{c^2} \end{bmatrix} = \left(\frac{3v^2}{r^2 c^2}\right) \cdot \begin{bmatrix} 1 & \frac{v_x}{c} & \frac{v_y}{c} & \frac{v_z}{c} \\ \frac{v_x}{c} & \frac{v_x^2}{c^2} & \frac{v_x v_y}{c^2} & \frac{v_x v_z}{c^2} \\ \frac{v_y}{c} & \frac{v_x v_y}{c^2} & \frac{v_y^2}{c^2} & \frac{v_y v_z}{c^2} \\ \frac{v_z}{c} & \frac{v_x v_z}{c^2} & \frac{v_y v_z}{c^2} & \frac{v_z^2}{c^2} \end{bmatrix}$$